\documentclass[aps,preprint,tightenlines,showpacs]{revtex4-1}
\usepackage{amssymb}

\usepackage{dcolumn}
\usepackage{bm}

\begin{document}

\title{Maxwell-affine gauge theory of gravity}
\author{O. Cebecio\u{g}lu$^1$ \footnote{%
E-mail: ocebecioglu@kocaeli.edu.tr} and S. Kibaro\u{g}lu$^1$ \footnote{%
E-mail: salihkibaroglu@gmail.com}}
\date{\today}

\begin{abstract}
Maxwell extension of affine algebra with additional tensorial generators is
given. Using the methods of nonlinear realizations, we found the
transformation rules for group parameters and corresponding generators.
Gauging the Maxwell-affine algebra we presented two possible invariant
actions for gravity: one is the first order and the other one is the second
order in affine curvature. We noticed that equations of motion for the
action, second order in affine curvature, lead to the generalized Bianchi
identities on the choice of appropriate coefficients for a particular
solution of the constraint equation.
\end{abstract}

\affiliation{$^1$Department of Physics, Kocaeli University, 41380 Kocaeli,
Turkey}

\pacs{02.20.Sv; 04.20.Fy; 11.15.-q; 02.40.-k}

\maketitle


\section{Introduction}

Following the pioneering work of Bacry \cite{BCR}, the idea of Maxwell
symmetry has been systematically studied by Schrader \cite{SCH}, and has
received much attention, starting again with Soroka \cite{SS1}, over the
past 10 years or so. The attention has focused mainly on two distinct
directions one with the dynamical particle realization both relativistic and
super symmetric \cite{BGKL1,BGKL2,BGKL3,FL} and the second direction is the
localization of the Maxwell symmetry in order to get extended gravity
theories \cite{AZKL,SS2}. The formulation of different types of Maxwell
gravities has already been studied in \cite{DG1,DG2,DG3,HA,AI}. In these
theories, Maxwell extension was applied to the Poincare, superPoincare and
(A)dS groups. With this motivation, we have presented the Maxwell-affine
gravity in a similar fashion by observing the result of \cite
{DG1,DG2,DG3,HA,AI}. We have already gone one more step beyond these groups
by adding dilatation into the scene \cite{OS} and now we are pushing even
more step beyond these taking into account the group of affine
transformations. The aim of this paper is to establish the framework of the
gauge theory of Maxwell-affine group, $\mathcal{MA}(4,R)$. This will be done
by constructing the noncentral extension of affine group. As we shall see,
the Maxwell-affine algebra is constructed from the translation generators $%
P_{a}$ and the noncentral extension $Z_{ab}$ together with the $gl(4,R)$
generators ${L^{a}}_{b}$.

The organization of the paper is as follows. In Sec. II, after reviewing
some properties of the affine algebra, we study its extensions as described
in \cite{BG1,BG2}. Applying the techniques of non-linear coset realization 
\cite{CWZ,CCWZ,SAS1,SAS2} to the Maxwell-affine group, the transformation
rules for generalized coordinates are found and explicit expression for the
generators of the Maxwell-affine algebra are given. In Sec.III, we present
two types of action for gravity based on gauged Maxwell-affine algebra, one
with first order and the other second order in affine curvature. We derive
and discuss the field equations for both actions. Finally, Sec.IV concludes
the paper.

\section{Affine Algebra and Its Tensor Extension}

The affine group, $A(4,R)$, is a group of all linear transformations in four
dimensional space\cite{BO}: 
\begin{equation}
x^{a^{\prime }}={\Lambda ^{a^{\prime }}}_{b}x^{b}+c^{a}
\end{equation}
and it can be written as a semi direct product of its homogeneous and
inhomogeneous parts namely the general linear group, $GL(4,R)$, and the
translational group $T^{4}$ parts respectively. The 20 generators of affine
transformation can be decomposed into the 4 translations $P_{a}$ and 16
general linear transformations ${L^{a}}_{b}$. Its Lie algebra is defined by
the commutation relations 
\begin{eqnarray}
\left[ {L^{a}}_{b},{L^{c}}_{d}\right] &=&i({\delta ^{c}}_{b}L{^{a}}_{d}-{%
\delta ^{a}}_{d}L{^{c}}_{b})  \nonumber \\
\left[ {L^{a}}_{b},P_{c}\right] &=&-i{\delta ^{a}}_{c}P_{b}  \nonumber \\
\left[ P_{a},P_{b}\right] &=&0
\end{eqnarray}
where the tensor indices $a,b$ take the values $(0,1,2,3)$. The elements of
the affine group are here represented by exponentials according to 
\begin{equation}
g(x,\widetilde{\omega })=e^{ix^{a}P_{a}}e^{i\widetilde{\omega }{^{b}}_{a}L{%
^{a}}_{b}}
\end{equation}
where $e^{i\widetilde{\omega }{^{b}}_{a}L{^{a}}_{b}}$ $\in GL(4,R)$ \cite
{LTT,TM}.

The Maxwell extension of the affine algebra can be constructed in complete
analogy to the Maxwell algebra obtanined by \cite{BG1, BG2}. To construct
Maxwell extension of affine group, we need to introduce Maurer-Cartan, MC,
form defined by 
\begin{equation}
\Omega =-ig^{-1}dg
\end{equation}
where $g$ is a general element of affine group. This left invariant 1-form
satisfies the Maurer-Cartan structure equation:

\begin{equation}
d\Omega +\frac{i}{2}\left[ \Omega ,\Omega \right] =0.
\end{equation}
For the affine case, we have 
\begin{equation}
d\Omega {^{a}}_{P}+\Omega {^{a}}_{Lb}\wedge \Omega {^{b}}_{P}=0
\end{equation}
\begin{equation}
d\Omega {^{a}}_{Lb}+\Omega {^{a}}_{Lc}\wedge \Omega {^{c}}_{Lb}=0.
\end{equation}
Freezing the GL(4,R) degrees of freedom ($L{^{a}}_{b}\rightarrow 0),$ the
most general closed invariant 2-form which cannot be written as the
differential of an invariant 1-form is of the form 
\begin{equation}
\Omega _{2}=f_{\left[ ab\right] }\Omega {^{a}}_{P}\wedge \Omega {^{b}}_{P}
\end{equation}
where the constant parameters $f_{\left[ ab\right] }$ is a second rank
antisymmetric tensor. Therefore, we find that the non-trivial 2-forms an
antisymmetric tensor representation of the $GL(4,R)$ group. The 1-form
potential associated to this 2-form is denoted by $Z_{ab}$ and satisfy he MC
structure equation 
\begin{equation}
d\Omega {^{ab}}_{Z}-\frac{1}{2}\Omega {^{a}}_{P}\wedge \Omega {^{b}}_{P}=0
\end{equation}
where the coefficient $(-\frac{1}{2})$ chosen for notational convenience.
When the general linear transformation included, the extended set of MC
1-forms satisfies the equations 
\begin{eqnarray}
d\Omega {^{a}}_{P}+\Omega {^{a}}_{Lb}\wedge \Omega {^{b}}_{P} &=&0  \nonumber
\\
d\Omega {^{a}}_{Lb}+\Omega {^{a}}_{Lc}\wedge \Omega {^{c}}_{Lb} &=&0 
\nonumber \\
d\Omega _{Z}^{ab}+\Omega {^{\lbrack a}}_{Lc}\wedge \Omega {^{cb]}}_{Z}-\frac{%
1}{2}\Omega _{P}^{a}\wedge \Omega _{P}^{b} &=&0  \label{MC}
\end{eqnarray}
where second term in third equation implies that$\ Z_{ab}$ generators
transform as a tensor with respect to $GL(4,R)$ transformations. Eqs.(\ref
{MC}) implies the 26 dimensional Maxwell-affine algebra with the following
non-zero commutation rules: 
\begin{eqnarray}
\left[ {L^{a}}_{b},{L^{c}}_{d}\right] &=&i({\delta ^{c}}_{b}L{^{a}}_{d}-{%
\delta ^{a}}_{d}L{^{c}}_{b})  \nonumber \\
\left[ {L^{a}}_{b},P_{c}\right] &=&-i{\delta ^{a}}_{c}P_{b}  \nonumber \\
\left[ {P}_{a},P_{b}\right] &=&iZ_{ab}  \nonumber \\
\left[ {L^{a}}_{b},Z_{cd}\right] &=&i({\delta ^{a}}_{d}Z_{bc}-{\delta ^{a}}%
_{c}Z_{bd})
\end{eqnarray}

To realize the group action on $\mathcal{M}\mathcal{A}(4,R)/{GL}(4,R)$, we
make use of the formula 
\begin{equation}
g(a,\varepsilon ,u)K(x,\theta )=K(x^{\prime },\theta ^{^{\prime }})h(%
\widetilde{\omega },g)
\end{equation}
which defines the non-linear group action, choosing exponential
parametrization for the coset 
\begin{equation}
K(x,\theta )=e^{ixp}e^{i\theta Z}
\end{equation}
where the variables $x^{a},\theta ^{ab}$ the coset parameters and 
\begin{equation}
h(\widetilde{\omega })=e^{i\widetilde{\omega }{^{b}}_{a}L{^{a}}_{b}}
\end{equation}
an element of stability subgroup ${GL}(4,R)$. It can be easily evaluated
through the use of the well known Baker-Hausdorff-Campell formula:$\ $%
\begin{equation}
e^{A}e^{B}=e^{A+B+\frac{1}{2}\left[ A,B\right] }
\end{equation}
which holds when $\left[ A,B\right] $ commutes with both $A\ $and$\ B.$ The
transformation laws of the coset space parameters under the infinitesimal
action of the $\mathcal{MA}(4,R)$ are 
\begin{eqnarray}
\delta x^{a} &=&a^{a}+u{^{a}}_{b}x^{b}  \nonumber \\
\delta \theta ^{ab} &=&\varepsilon ^{ab}-\frac{1}{4}a^{[a}x^{b]}+u{^{[a}}%
_{c}\theta ^{cb]}  \nonumber \\
\widetilde{\omega }{^{a}}_{b} &=&u{^{a}}_{b}
\end{eqnarray}
where antisymmetrization is defined by $x^{[a}y^{b]}=x^{a}y^{b}-x^{b}y^{a}$
and the corresponding generators are 
\begin{eqnarray}
P_{a} &=&i(\partial _{a}-\frac{1}{2}x^{b}\partial _{ab})  \nonumber \\
Z_{ab} &=&i\partial _{ab}  \nonumber \\
L{^{a}}_{b} &=&i(x^{a}\partial _{b}+2\theta {^{ac}}\partial _{bc})
\end{eqnarray}
where $\partial _{a}=\frac{\partial }{\partial x^{a}}$, $\partial _{ab}=%
\frac{\partial }{\partial \theta ^{ab}},$ and one can check that these
generators fulfill the Maxwell-affine algebra and verify self-consistency of
Jacobi identities.

\section{Gauging the Maxwell-affine algebra}

\bigskip The gauge theories of affine gravity treated in \cite{HLN,LORD}.
The nonlinear \ gauge theories of gravity on the basis of the affine group $%
A(4,R)$ as the principal group was used in \cite{BO,LTT,TM}. More complete
references on affine gauge theory and the metric affine gravity up to 1995
can be found in \cite{HMMN}. \ To gauge affine algebra we introduce the
1-form potential $\mathcal{A}$ with the values in Lie algebra of the $%
\mathcal{MA}(4,R)$ group, defined by 
\begin{equation}
\mathcal{A}=e^{a}P_{a}+B^{ab}Z_{ab}+\widetilde{\omega }{^{b}}_{a}L{^{a}}_{b}
\end{equation}
where $e^{a}$ $=e_{\mu }^{a}dx^{\mu }$, $B^{ab}={B^{ab}}_{\mu }dx^{\mu }$
and $\widetilde{{\omega }}{^{a}}_{b}=\widetilde{{\omega }}{^{a}}_{\mu
b}dx^{\mu }$ are vector fields with respect to the space-time index ${\mu }$%
. The variation of these fields under infinitesimal gauge transformations in
tangent space is given by 
\begin{equation}
\delta \mathcal{A}=-d\zeta -i\left[ \mathcal{A},\zeta \right]
\label{delta_amu}
\end{equation}
with the gauge generator 
\begin{equation}
\zeta \left( x\right) =y^{a}\left( x\right) P_{a}+\varphi
^{ab}(x)Z_{ab}+\lambda {^{b}}_{a}(x)L{^{a}}_{b}
\end{equation}
where $y^{a}\left( x\right) $ are space-time translations, $\varphi ^{ab}(x)$
are translations in tensorial space, and $\lambda {^{b}}_{a}(x)$ are the
general linear transformation parameters respectively. The transformation
properties of the 26 gauge fields under infinitesimal $\mathcal{MA}(4,R)$
are 
\begin{eqnarray}
\delta e^{a} &=&-dy^{a}-\widetilde{{\omega }}{^{a}}_{b}y^{b}+\lambda {^{a}}%
_{b}\ e^{b}  \nonumber \\
\delta B^{ab} &=&-d\varphi ^{ab}-\widetilde{{\omega }}{^{[a}}_{c}\varphi {%
^{cb]}}+{\lambda ^{\lbrack a}}_{c}B{^{cb]}}+\frac{1}{2}e^{[a}y^{b]} 
\nonumber \\
\delta \widetilde{{\omega }}{^{a}}_{b} &=&-d\lambda {^{a}}_{b}-\widetilde{{%
\omega }}{^{a}}_{c}\lambda {^{c}}_{b}+{\lambda ^{a}}_{c}\widetilde{{\omega }}%
{^{c}}_{b}  \label{delta_e}
\end{eqnarray}
The curvature 2-form $\mathcal{\digamma }$ is given by the structure
equation 
\begin{equation}
\ \mathcal{\digamma }\ =d\mathcal{A}+\frac{i}{2}\left[ \mathcal{A},\mathcal{A%
}\right]
\end{equation}
whence writing 
\begin{equation}
\mathcal{\digamma }=F^{a}P_{a}+F^{ab}Z_{ab}+\widetilde{R}{^{b}}_{a}L{^{a}}%
_{b}  \label{f1}
\end{equation}
we find 
\begin{eqnarray}
F^{a} &=&de^{a}+\widetilde{{\omega }}{^{a}}_{b}\wedge e^{b}  \nonumber \\
F^{ab} &=&dB^{ab}+\widetilde{{\omega }}{^{[a}}_{c}\wedge B^{cb]}-\frac{1}{2}%
e^{a}\wedge e^{b}  \nonumber \\
\widetilde{R}{^{a}}_{b} &=&d\widetilde{{\omega }}{^{a}}_{b}+\widetilde{{%
\omega }}{^{a}}_{c}\wedge \widetilde{{\omega }}{^{c}}_{b}
\end{eqnarray}
These are the general-affine torsion, a new curvature 2-form for tensor
generator $Z_{ab}$, and the general-affine curvature 2-form respectively.
Under an infinitesimal gauge transformation with parameters $\zeta $, the
curvature 2-form $\ \mathcal{\digamma }$ transform as 
\begin{equation}
\delta \mathcal{\digamma }=i\left[ \zeta ,\mathcal{\digamma }\right]
\end{equation}
and hence one gets 
\begin{eqnarray}
\delta F^{a} &=&-\widetilde{R}{^{a}}_{b}y^{b}+\lambda {^{a}}_{b}F^{b} 
\nonumber \\
\delta F^{ab} &=&-\widetilde{R}{^{[a}}_{c}\varphi ^{cb]}+\lambda {^{\lbrack
a}}_{c}F^{cb]}-\frac{1}{2}y^{[a}F^{b]}  \nonumber \\
\delta \widetilde{R}{^{a}}_{b} &=&\lambda {^{a}}_{c}\widetilde{R}{^{c}}_{b}-%
\widetilde{R}{^{a}}_{c}\lambda {^{c}}_{b}.  \label{delta_F}
\end{eqnarray}

Having found the transformation of the gauge fields and the curvatures, we
are ready to look for invariant lagrangians under these transformations. We
assume that the generalized gravitational Lagrangian 4-form to be gauge
invariant under the homogeneous local $GL(4,R)$\ subgroup only. To construct
gauge invariant Lagrangian 4-form first order in affine curvature, we start
from the Euler density in four dimensions and substitute each curvature by a
concircular curvature\cite{MZ}. 
\begin{equation}
R{^{ab}}\rightarrow \overline{R}{^{ab}(\beta )}=R{^{ab}}-\beta e{^{a}\wedge }%
e^{b}
\end{equation}
where the last term recalls the contribution to Lorentz curvature $R{^{ab}}$
in (A)dS gravity, enters through the new gauge field strenght $F{^{ab}}$. In
the light of this substitution we combine the curvatures $\widetilde{R}{^{ab}%
}$ and $F{^{ab}}$ into a single curvature as follows 
\begin{equation}
\widetilde{R}{^{ab}}\rightarrow \mathcal{J}{^{ab}(\mu )}=\widetilde{R}{^{a}}%
_{e}g{^{eb}}-\mu F{^{ab}}
\end{equation}
where we also introduce an extra field, the premetric symmetric tensor field 
$g{^{ab}}$, its infinitesimal transformation under local $GL(4,R)$ is 
\begin{equation}
\delta g{^{ab}}=\lambda {^{a}}_{c}g{^{cb}}+\lambda {^{b}}_{c}g{^{ac}.}
\end{equation}
The reason for introducing this premetric symmetric field stems from
non-flatness of group space in ordinary affine gauge theory in contrast to
the local Lorentz group, the space defined by this group is flat and
characterized by flat Minkowski metric \cite{SO}. After these preliminaries
we start from the following gauge invariant Yang-Mills type action 
\begin{equation}
S_{1}=\frac{1}{2\varkappa }\int \mathcal{J}\wedge ^{\ast }\mathcal{J}=\frac{1%
}{4\varkappa }\int \eta _{abcd}\mathcal{J}^{ab}\wedge \mathcal{J}^{cd}
\label{action1}
\end{equation}
where \ $\eta _{abcd}=\sqrt{-g}$ $\varepsilon _{abcd}.$ Writing out $%
\mathcal{J}{^{ab}}$ in terms of the curvatures $\widetilde{R}{^{a}}_{e}g{%
^{eb}}$ and $F{^{ab}}$, the action becomes

\begin{equation}
S_{1}=\frac{1}{4\varkappa }\int \eta _{abcd}\left( \widetilde{R}{^{a}}_{e}g{%
^{eb}}\wedge \widetilde{R}{^{c}}_{f}g{^{fd}}-2\mu \widetilde{R}{^{a}}_{e}g{%
^{eb}}\wedge F^{cd}+\mu ^{2}F^{ab}\wedge F^{cd}\right) .  \label{action2}
\end{equation}
The second term of this Maxwell affine action contains the direct
generalization of the Einstein Hilbert action to the affine case \cite{LEC}.

One discovers that variation with respect to $\widetilde{\omega }{^{b}}_{a}$
gives the generalized torsion 
\begin{equation}
\mathcal{D}\left( g^{ac}{}^{\ast }\mathcal{J}_{bc}\right) -2\mu B^{ac}\wedge 
\mathcal{J}_{cb}=0
\end{equation}
while variation with respect to the vierbein field $e^{a}$ gives the field
equation 
\begin{equation}
^{\ast }\mathcal{J}_{ab}\wedge e{^{b}}=0.  \label{Eins}
\end{equation}
By a variation with respect to premetric field $g{^{ab}}$, we get 
\begin{equation}
\widetilde{R}{^{c}}_{(a}\wedge ^{\ast }\mathcal{J}_{cb)}-\frac{1}{2}g_{ab}%
\mathcal{J}{^{cd}}\wedge ^{\ast }\mathcal{J}_{cd}=0  \label{premetric}
\end{equation}
where symmetrization is defined by $A_{(a}\wedge B_{b)}=A_{a}\wedge
B_{b}+A_{b}\wedge B_{a}$. Finally, variation with respect to $B^{ab}$ leads
to 
\begin{equation}
\mathcal{D}^{\ast }\mathcal{J}_{ab}=0
\end{equation}
where $\mathcal{D}=d+\widetilde{\omega }$ is the general linear exterior
covariant derivative. We conclude our discussion by emphasizing that Eq.(\ref
{premetric}) is the generalization of the equation given in \cite{JS1}.
Remaining three equations of motion are local $GL(4,R)$ modified version of
Azcarraga's equations of motion \cite{AZKL}. Thus, the field equations of
Einstein-Cartan like theory are recovered.

To construct gauge invariant action whose Lagrangian 4-form, second order in
curvatures, we start from the Pontryagin densities in 4 dimensions and
substitute each product of two curvatures \cite{MZ} according to 
\begin{equation}
R{^{a}}_{b}\wedge R{^{b}}_{a}\rightarrow \overline{R}{^{a}}_{b}\left( \beta
\right) \wedge \overline{R}{^{b}}_{a}\left( \beta \right) -\gamma
F^{a}\wedge F_{a}
\end{equation}
where $\beta$ and $\gamma$ are scale constants and $R{^{a}}_{b}$ is a
Lorentz curvature. This time we chose our curvature as $\mathcal{Y}{^{a}}%
_{b}=\widetilde{R}{^{a}}_{b}-\mu F{^{a}}_{b}$ without any reference to
premetric tensor field because in this form it transforms as a nice tensor
under local $GL(4,R)$ which makes it possible to construct a gauge invariant
action in the form of

\begin{equation}
S_{2}=\frac{1}{2\varkappa }\int \mathcal{Y}{^{a}}_{b}\wedge \mathcal{Y}{^{b}}%
_{a}+\frac{1}{\rho }\int F^{a}\wedge F_{a}  \label{action}
\end{equation}
where$\mathcal{\ }$ $\varkappa $ and $\rho =-\frac{2\varkappa }{\gamma }$
are coupling constants. By construction the action Eq.(\ref{action}) is
manifestly diffeomorphism invariant and possess local $GL(4,R)$ invariance.
The field equations follow from the variation of the action. From the
variation of Eq.(\ref{action}) with respect to $\widetilde{\omega }$ we get
the following equation for the generalized torsion tensor 
\begin{equation}
\mathcal{DY}{^{a}}_{b}-\mu \left[ B,\mathcal{Y}\right] {^{a}}_{b}+\frac{%
\varkappa }{\rho }\left[ F,e\right] {^{a}}_{b}=0.  \label{B1}
\end{equation}
where $\left[ B,\mathcal{Y}\right] {^{a}}_{b}$ denotes taking the wedge
product on the form part and the commutator on the Lie algebra part, i.e. $%
\left[ B,\mathcal{Y}\right] {^{a}}_{b}$ $=(B{^{a}}_{c}\wedge \mathcal{Y}{^{c}%
}_{b}-\mathcal{Y}{^{a}}_{c}\wedge B{^{c}}_{b})$, similarly we have $\left[
F,e\right] {^{a}}_{b}=(F{^{a}}\wedge e_{b}-e{^{a}}\wedge F_{b}).$
Furthermore, the $e$ variation of the action gives following two equations 
\begin{eqnarray}
\mathcal{Y}{^{a}}_{b}\wedge e^{b}-\left( \frac{2\varkappa }{\mu \rho }%
\right) \mathcal{D}F^{a} &=&0  \nonumber \\
e_{b}\wedge \mathcal{Y}{^{b}}_{a}+\left( \frac{2\varkappa }{\mu \rho }%
\right) \mathcal{D}F_{a} &=&0.  \label{B2}
\end{eqnarray}
and finally 
\begin{equation}
\mathcal{DY}{^{a}}_{b}=0  \label{B3}
\end{equation}
obtained by varying with respect to the $B$. Substituting Eq.(\ref{B3}) into
Eq.(\ref{B1}) and taking exterior covariant derivative and making use of Eq.(%
\ref{B2}) leads the following constraint equation 
\begin{equation}
F{^{a}}_{c}\wedge \widetilde{R}{^{c}}_{b}-\widetilde{R}{^{a}}_{c}\wedge F{%
^{c}}_{b}=0.  \label{constraint}
\end{equation}
From \bigskip special solution of Eq.(\ref{constraint}) 
\begin{equation}
\widetilde{R}{^{c}}_{b}=-\mu F{^{c}}_{b}
\end{equation}
we get 
\begin{equation}
\mathcal{Y}{^{a}}_{b}=-2\mu F{^{a}}_{b}=2\widetilde{R}{^{a}}_{b}
\end{equation}
If we insert this solution to the equations of motion, we arrived the result
that equations of motion are the generalized Bianchi identities as the $%
\left( \frac{\varkappa }{\mu \rho }\right) \rightarrow 1$ limit, 
\begin{eqnarray}
\mathcal{D}F^{ab} &=&\widetilde{R}{^{[a}}_{c}\wedge B^{cb]}-\frac{1}{2}F{%
^{[a}}\wedge e^{b]}=0  \nonumber \\
\mathcal{D}F^{a} &=&\widetilde{R}{^{a}}_{b}\wedge e^{b}  \nonumber \\
\mathcal{D}\widetilde{R}{^{a}}_{b} &=&0.
\end{eqnarray}

\bigskip For completenes let us give the conservation laws which follow from
the invariance of the action under the local $\mathcal{MA}(4,R)$ symmetry.
Under diffeomorphism, the variation of the Lagrangian is given by its Lie
derivative $l_{\xi }\mathcal{L}=di_{\xi }\mathcal{L}+i_{\xi }d\mathcal{L}$
along $\xi $. Now $d\mathcal{L}=0$ because $\mathcal{L}$ is a top form and
the first term is a total divergence which can be ignored as a surface
integral then we have 
\begin{equation}
\ \delta S_{diff}=\int l_{\xi }\mathcal{L}=0.
\end{equation}
In order to show the diffeomorphism invariance of the action explicitly, one
substitudes the transformation rules Eq.(\ref{delta_e}) to the following
action integral 
\begin{equation}
\ \delta S=\int \delta e^{a}\wedge E_{a}+\delta B^{ab}\wedge V_{ab}+\delta 
\widetilde{\omega }{^{b}}_{a}\wedge C{^{a}}_{b}
\end{equation}
then one gets the following conservation rules 
\begin{eqnarray}
\mathcal{D}E_{a}+V_{ab}\wedge e^{b} &=&0  \nonumber \\
\mathcal{D}V_{ab} &=&0  \nonumber \\
\mathcal{D}C{^{a}}_{b}+2B{^{ac}}\wedge V_{cb}+e^{a}\wedge E_{b} &=&0
\end{eqnarray}
where \bigskip $E_{a}$, $C{^{a}}_{b}$, and $V_{ab}$ are Einstein ,Cartan and
Maxwell 3-forms respectively. They can be found from actions Eq.(\ref
{action2}) and Eq.(\ref{action}) for the respective Lagrangians.

\section{Conclusion}

In the present paper we enlarged the general affine algebra by using an
antisymmetric tensor generator and constructed a non-linear realization of
the Maxwell-affine group on its coset space with respect to the general
linear group. We have shown two new set of field equations within the
framework of the Maxwell affine gauge theory of gravity starting from Euler
density, we formed an action which is linear in affine curvature leading to
the generalized Einstein-Hilbert action and thus generalizing the results of 
\cite{JS1},  and we have also constructed an action second order in
curvatures by making use of Pontryagin density which in turn leads to the
generalized Bianchi identities. \ 

\section*{Acknowledgements}

The authors wish to thank Abdurrahman Andi\c{c} and Mustafa Erkovan for
helpful discussions and useful remarks.


\begin{thebibliography}{99}
\bibitem{BCR}  H. Bacry, P. Combe, and J.L. Richard, Nuovo Cimento \textbf{67%
}, 267-299 (1970).

\bibitem{SCH}  R. Schrader, 
Fortschritte der Physik \textbf{20}, 701 (1972).

\bibitem{SS1}  D.V. Soroka and V.A. Soroka, 
Phys. Lett. B \textbf{607}, 302-305 (2005).

\bibitem{BGKL1}  J. Gomis, K. Kamimura, and J. Lukierski 
JHEP \textbf{08}, 39 (2009).

\bibitem{BGKL2}  S. Bonanos, J. Gomis, K. Kamimura, and J. Lukierski, 
Phys. Rev. Lett. \textbf{104}, 090401 (2010).

\bibitem{BGKL3}  S. Bonanos, J. Gomis, K. Kamimura, and J. Lukierski 
J. Math. Phys. \textbf{51}, 102301 (2010).

\bibitem{FL}  S. Fedoruk and J. Lukierski, 
JHEP \textbf{02}, 128 (2013).

\bibitem{AZKL}  J.A. de Azcarraga, K. Kamimura, and J. Lukierski, 
Phys. Rev. D \textbf{83}, 124036, (2011). 

\bibitem{SS2}  D.V. Soroka and V.A. Soroka, 
Phys. Lett. B \textbf{707}, 160-162 (2012). 

\bibitem{DG1}  R. Durka, Kowalski-Glikman, and M. Szczachor, 
Mod. Phys. Lett. A \textbf{26}, 2689-2696 (2011).

\bibitem{DG2}  R. Durka, Kowalski-Glikman, and M. Szczachor, 
Mod. Phys. Lett. A \textbf{27}, 1250023 (2012).

\bibitem{DG3}  R. Durka and Kowalski-Glikman, 
arXiv:1110.6812v1 [hep-th].

\bibitem{HA}  S. Hoseinzadeh and A. Rezaei-Aghdam, 
Phys. Rev. D \textbf{90}, 084008 (2014).

\bibitem{AI}  J. A. de Azcarraga and J. M. Izquierdo, 
Nucl. Phys. B \textbf{885}, 34-45 (2014). 

\bibitem{OS}  O. Cebecioglu and S. Kibaroglu, 
Phys. Rev. D \textbf{90}, 084053 (2014).

\bibitem{BG1}  S. Bonanos and J. Gomis, 
J. Phys. A: Math. Theor. \textbf{42}, 145206 (2009).

\bibitem{BG2}  S. Bonanos and J. Gomis, 
J. Phys. A: Math. Theor. \textbf{43}, 015201 (2010). 

\bibitem{CWZ}  S. Coleman, J. Wess, and B. Zumino, Phys. Rev. \textbf{177},
2239 (1969).

\bibitem{CCWZ}  C. Callan, S. Coleman, J. Wess, and B. Zumino, Phys. Rev. 
\textbf{177}, 2247 (1969).

\bibitem{SAS1}  A. Salam and J. Strathdee, Phys. Rev. \textbf{184}, 1750
(1969).

\bibitem{SAS2}  A. Salam and J. Strathdee, Phys. Rev. \textbf{184}, 1760
(1969).

\bibitem{BO}  A.B. Borisov and V.I. Ogievetskii, 
Theor. Math. Phys. \textbf{21}, 1179-1188 (1974).

\bibitem{LTT}  A. Lopez-Pinto, A. Tiemblo, and R. Tresguerres, 
Class.Quant.Grav. \textbf{12}, 1503-1516 (1995). 

\bibitem{TM}  R. Tresguerres and E.W. Mielke, 
Phys. Rev. D \textbf{62}, 044004 (2000).

\bibitem{HLN}  F. W. Hehl, E. A. Lord, and Y. Ne'eman, 
Phys. Rev. D \textbf{17}, 428-433 (1978).

\bibitem{LORD}  E. Lord, 
Phys.Lett. A \textbf{65}, 1-4 (1978). 

\bibitem{HMMN}  F.W. Hehl, J.D. McCrea, E.W. Mielke, and Y. Ne'eman, 
Phys. Rept. \textbf{258}, 1-171 (1995). 

\bibitem{MZ}  A. Mardones and J. Zanelli, 
Class. Quant. Grav. \textbf{8}, 1545-1558 (1991). 

\bibitem{SO}  R. F. Sobreiro and V. J. Vasquez Otoya, 
J. Geom. Phys. \textbf{61}, 137-150 (2011) 

\bibitem{LEC}  M. Leclerc, 
Annals Phys. \textbf{321}, 708-743 (2006). 

\bibitem{JS1}  B. Julia and S. Silva, 
Class. Quant. Grav. \textbf{15}, 2173-2215 (1998). 
\end{thebibliography}

\end{document}